\documentclass[aps,prl,showpacs,notitlepage,twocolumn,superscriptaddress,nofootinbib,preprintnumbers]{revtex4-2}
\usepackage{wrapfig}
\usepackage{bbm}
\usepackage{url}
\usepackage{mathrsfs}
\usepackage{epsfig}
\usepackage[utf8]{inputenc}
\usepackage{soul,xcolor}
\usepackage{graphicx}
\usepackage{amsfonts}
\usepackage{amsthm}
\usepackage[figuresright]{rotating}
\usepackage{amssymb}
\usepackage{amsmath}
\usepackage{dcolumn}
\usepackage{physics}
\usepackage{float}
\usepackage{bm}
\usepackage{physics}
\usepackage{verbatim}
\usepackage{pifont}  
\newcommand{\cmark}{\checkmark}
\newcommand{\xmark}{\ding{55}}
\usepackage{braket}
\usepackage[normalem]{ulem}
\usepackage[ruled,vlined,linesnumbered]{algorithm2e}
\usepackage{setspace}
\usepackage{lipsum}
\usepackage[colorlinks,linkcolor=blue,anchorcolor=blue,citecolor=blue,urlcolor=blue]{hyperref}
\usepackage{graphicx}
\usepackage{amsmath, nccmath}
\usepackage{amsfonts}
\usepackage{amssymb}
\usepackage{esvect}
\usepackage{amsthm}
\usepackage{mathdots}
\usepackage[caption=false]{subfig}
\usepackage{mathtools}
\usepackage{tensor}
\usepackage{mathrsfs}
\usepackage{comment}
\usepackage{bbm}
\usepackage{graphicx}
\usepackage{ragged2e}
\usepackage{color}
\usepackage{wrapfig}
\usepackage{color}
\usepackage{graphicx}
\usepackage{tikz}
\usepackage{dsfont}
\usepackage{mathbbol}
\usepackage{amssymb} 
\usepackage{simpler-wick}

\DeclareSymbolFontAlphabet{\amsmathbb}{AMSb}
\usetikzlibrary{shapes.misc}
\tikzset{cross/.style={cross out, draw=black, minimum size=2*(#1-\pgflinewidth), inner sep=0pt, outer sep=0pt},
cross/.default={3pt}}
\usetikzlibrary{arrows,shapes,positioning}
\usetikzlibrary{decorations.markings}
\usepackage[rightcaption]{sidecap}
\tikzstyle arrowstyle=[scale=1]
\tikzstyle directed=[postaction={decorate,decoration={markings,
    mark=at position .65 with {\arrow[arrowstyle]{stealth}}}}]
\tikzstyle reverse directed=[postaction={decorate,decoration={markings,
    mark=at position .65 with {\arrowreversed[arrowstyle]{stealth};}}}]
\usetikzlibrary{positioning}

\newcommand{\bea}{\begin{eqnarray}}
\newcommand{\eea}{\end{eqnarray}}
\newcommand{\be}{\begin{eqnarray}}
\newcommand{\ee}{\end{eqnarray}}
\newcommand{\bma}{\begin{matrix}}
\newcommand{\ema}{\cr\end{matrix}}

\newcommand{\U}{ {\text {U}}}

\newcommand{\integers}{ {\amsmathbb Z} }

\usepackage{ragged2e}
\usepackage{etoolbox}

\usepackage{ragged2e}
\usepackage{etoolbox}
\makeatletter
\patchcmd{\@makecaption}
  {\centering}
  {\justifying}
  {}
  {}
\makeatother

\def\cB{{\cal B}}

\def\cH{{\cal H}}

\def\cQ{{\cal Q}}

\def\cW{{\cal W}}

\def\half{{1\over 2}}

\def\a{\alpha}
\def\b{\beta}
\def\g{\gamma}
\def\d{\delta}

\def\g{\gamma}

\def\sfU{{\sf U}}

\def\sfE{{\sf E}}
\def\sfB{{\sf B}}
\def\sfG{{\sf G}}

\def\no{\nonumber}

\definecolor{darkred}{rgb}{0.8,0.1,0.1}
\hypersetup{colorlinks=true, linkcolor=darkred, citecolor=blue, linktoc=page}

\newcommand{\nc}{\newcommand}
\nc{\rnc}{\renewcommand} 

\rnc{\a}{\alpha}
\rnc{\b}{\beta}
\rnc{\d}{\delta}
\nc{\e}{\epsilon}
\nc{\z}{\zeta}
\nc{\m}{\mu}
\nc{\n}{\nu}
\rnc{\r}{\rho}
\rnc{\k}{\kappa}
\rnc{\l}{\lambda}
\nc{\s}{\sigma}
\rnc{\t}{\tau}
\nc{\w}{\omega}
\nc{\x}{\chi}
\nc{\F}{\Phi}
\rnc{\L}{\Lambda}

\nc{\pd}{\partial}

\makeatletter
\newcommand{\vast}{\bBigg@{4}}
\newcommand{\Vast}{\bBigg@{5}}
\makeatother

\begin{document}

\title{A Path to Quantum Simulations of Topological Phases:\\
(2+1)D Quantum Electrodynamics with Wilson Fermions}

\author{Sriram Bharadwaj}
\affiliation{
Mani L. Bhaumik Institute for Theoretical Physics, Department of Physics and Astronomy,\\ University of California, Los Angeles, CA 90095, USA
}
\author{Emil Rosanowski}
\affiliation{Transdisciplinary Research Area ``Building Blocks of Matter and Fundamental Interactions'' (TRA Matter) and Helmholtz Institute for Radiation and Nuclear Physics (HISKP), University of Bonn, Nussallee 14-16, 53115 Bonn, Germany}
\author{Simran Singh}
\affiliation{Transdisciplinary Research Area ``Building Blocks of Matter and Fundamental Interactions'' (TRA Matter) and Helmholtz Institute for Radiation and Nuclear Physics (HISKP), University of Bonn, Nussallee 14-16, 53115 Bonn, Germany}
\author{Alice Di Tucci}
\affiliation{Deutsches Elektronen-Synchrotron DESY, Platanenallee 6, 15738 Zeuthen, Germany}
\author{Changnan Peng}
\affiliation{Department of Physics, Massachusetts Institute of Technology, Cambridge, Massachusetts 02139, USA}
\author{Karl Jansen}
\affiliation{Deutsches Elektronen-Synchrotron DESY, Platanenallee 6, 15738 Zeuthen, Germany}
\affiliation{
 Computation-Based Science and Technology Research Center, The Cyprus Institute, 20 Kavafi Street,
2121 Nicosia, Cyprus
}
\author{Lena Funcke}
\affiliation{Transdisciplinary Research Area ``Building Blocks of Matter and Fundamental Interactions'' (TRA Matter) and Helmholtz Institute for Radiation and Nuclear Physics (HISKP), University of Bonn, Nussallee 14-16, 53115 Bonn, Germany}
\author{Di Luo}
\thanks{diluo1000@gmail.com}
\affiliation{Department of Physics, Tsinghua University, Beijing 100084, China}
\affiliation{Institute of Advanced Study, Tsinghua University, Beijing 100084, China}
\affiliation{Department of Electrical and Computer Engineering, University of California, Los Angeles, CA 90095, USA}

\begin{abstract}
Quantum simulation offers a powerful approach to studying quantum field theories, particularly (2+1)D quantum electrodynamics (QED$_3$) with Wilson fermions, which hosts a rich landscape of physical phenomena. A key challenge in lattice formulations is the proper realization of topological phases and the Chern-Simons terms, where fermion discretization plays a crucial role. In this work, we highlight the differences between staggered and Wilson fermions coupled to $\text{U}(1)$ gauge fields in the Hamiltonian formulation. We analyze why staggered fermions fail to induce (2+1)D topological phases, while Wilson fermions admit a variety of topological phases including Chern insulator and quantum spin Hall phases. Additionally, we uncover a rich phase diagram for the two-flavor Wilson fermion model in the presence of a chemical potential. Our findings resolve existing ambiguities in Hamiltonian formulations and provide a theoretical foundation for future quantum simulations of lattice field theories with topological phases. We further outline connections to experimental platforms, offering guidance for implementations on near-term quantum computing architectures. A complementary presentation of the analytical calculations, the identification of robust topological structure and response, and extensive numerical results is contained in a joint submission \cite{PRD}.
\end{abstract}
\maketitle

\emph{Introduction}---. Quantum simulation has emerged as a powerful tool for studying quantum field theory and high-energy physics, enabling the exploration of strongly interacting systems beyond the reach of classical computation~\cite{Bauer:2022hpo, DiMeglio:2023nsa}. In particular, quantum simulation provides a promising avenue to study real-time dynamics in quantum field theory, circumventing the sign problem that plagues classical Monte Carlo methods~\cite{Troyer:2004ge}. Furthermore, it enables the direct realization of lattice gauge theories in controllable quantum hardware, offering new insights into nonperturbative phenomena in high-energy physics~\cite{Banuls:2019bmf, Funcke:2023jbq}. A particularly exciting frontier is the simulation of (2+1)D Quantum Electrodynamics (QED$_3$), which exhibits rich emergent phenomena such as confinement, chiral symmetry breaking, and potential realizations of exotic topological phases~\cite{Pisarski:1984dj, Polyakov:1975rs, Polyakov:1976fu, peng2024hamiltonian}. Recent advancements in quantum computing platforms~\cite{Paulson:2020zjd,Haase:2020kaj,Crippa:2024hso, Crippa:2024cqr, Cochran:2024rwe,gonzalez2024observation, Meth:2023wzd, Ciavarella:2024fzw}, together with tensor network methods~\cite{Tagliacozzo:2014bta,Zapp:2017fcr,Felser:2019xyv,Xu:2025ean} and neural network representations of quantum states~\cite{luo2021gauge_equ,luo2021gauge_inv,luo2022gauge,chen2022simulating}, have opened new possibilities for simulating QED$_3$ on discrete lattices. However, constructing Hamiltonian formulations that accurately capture gauge invariance, fermionic dynamics, and topological effects remains a significant challenge. 

A key open question in the lattice formulation of QED$_3$ is the realization of topological phases and the Chern-Simons term, which plays a crucial role in topological gauge theories and fractional quantum Hall physics. The lattice Hamiltonian approach often introduces ambiguities, particularly in the formulation of fermions. Staggered fermions, a commonly used discretization scheme to reduce fermion doubling, have been widely employed in lattice gauge theory, but their ability to capture topological effects remains unclear. In particular, there is confusion about whether staggered fermions can host nontrivial Chern numbers and whether they effectively reproduce continuum topological phases. A deeper understanding of these issues is essential for designing quantum simulation experiments capable of probing topological effects in gauge theories.
In the Lagrangian formulation of lattice gauge theories, Wilson fermions coupled to $\U(1)$ gauge fields host a rich set of infrared topological phases \cite{Kaplan:1992bt,Golterman:1992ub,Jansen:1992tw,Jansen:1992yj}. Beyond the plane-wave approximation \cite{Jansen:1992yj}, a Hamiltonian analysis of topological phases has been carried out for free Dirac/Wilson fermions without coupling to gauge fields that satisfy Gauss' law \cite{Tirrito:2018bui}, and also for (3+1)D axionic models \cite{Bermudez:2010da}. 

\begin{figure*}[t]
    
    \centering
    \includegraphics[width=0.85\linewidth]{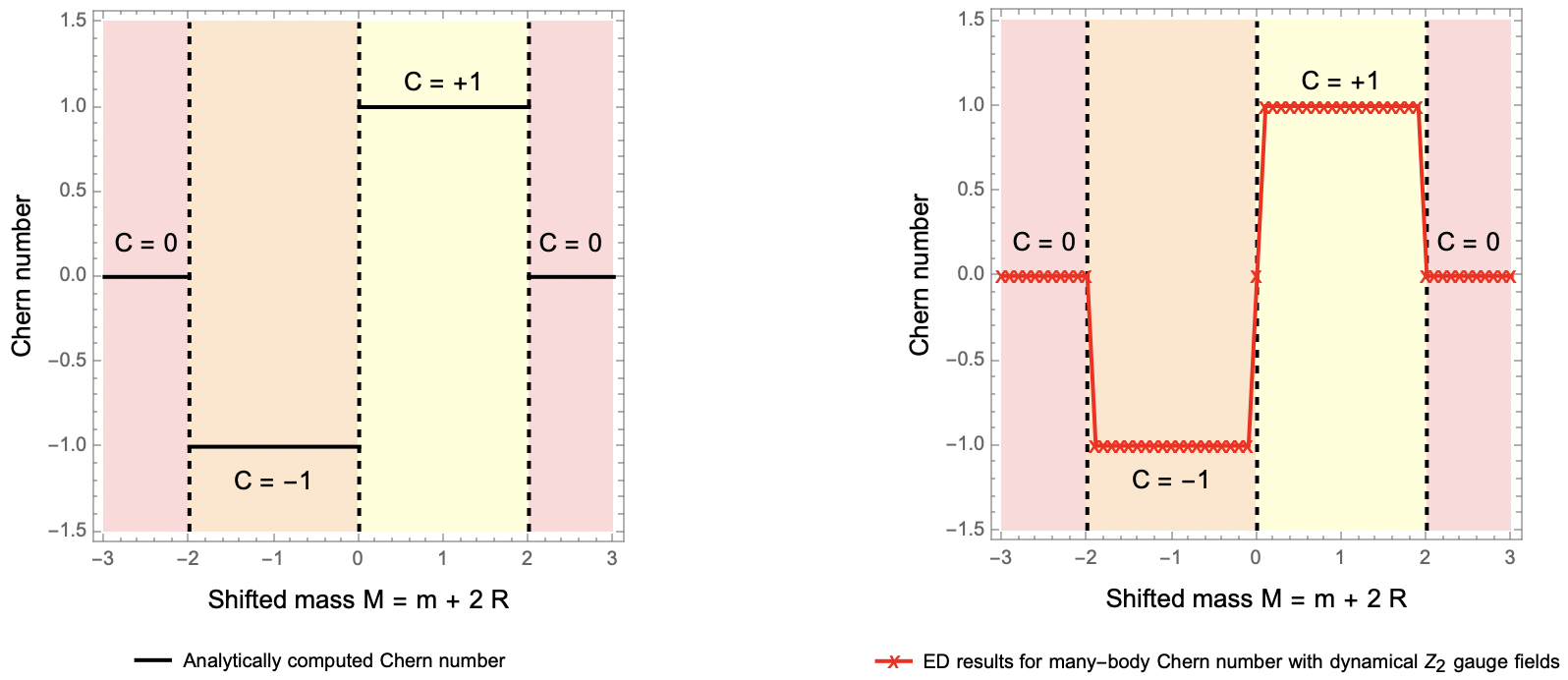}
    \caption{
    The analytically computed Chern number $C$ (left) vs.\ the shifted mass $M=m+2R$ for a $\U(1)$ gauge group and arbitrary system size $L$, and the numerically obtained many-body Chern number (right) on a $2\times 2$ lattice with $\integers_2$ gauge fields for $e^2=0.01$. The pink regions are trivial insulator phases, the orange one has $C = -1$, while the yellow one has $C = +1$. This figure demonstrates the robustness of topological phases even under severe gauge-group and lattice truncations, making them ideal targets for quantum simulations. Derivations of the analytical and many-body Chern numbers are relegated to Ref~\cite{PRD}.\label{fig:ChNo}}
    
\end{figure*}

In this work,\footnote{This work is part of a joint submission. A companion paper, Ref.~\cite{PRD}, presents complementary results and analyses related to the findings reported here.} we highlight the role of fermion discretizations in the emergence of (2+1)D topological phases in U(1) Hamiltonian lattice gauge theory. First, we elucidate why staggered fermions fail to host topological phases by emphasizing the role of broken time-reversal symmetry. Second, we present the infrared phase diagrams of weakly-coupled QED$_3$, with both one- and two-species of Wilson fermions at finite-density, which naturally exhibits a variety of topological phenomena, including Chern insulator and Quantum Spin Hall phases. Our findings provide a theoretical foundation for constructing lattice gauge models that accurately encode topological effects in quantum simulations. Finally, we outline key directions for future quantum simulations, establishing connections between lattice Hamiltonian formulations and experimental implementations on near-term quantum computers. Our results offer a crucial step toward realizing quantum simulations of topological phases of gauge theories.

\begin{figure*}
\centering
\includegraphics[width=0.35\linewidth]{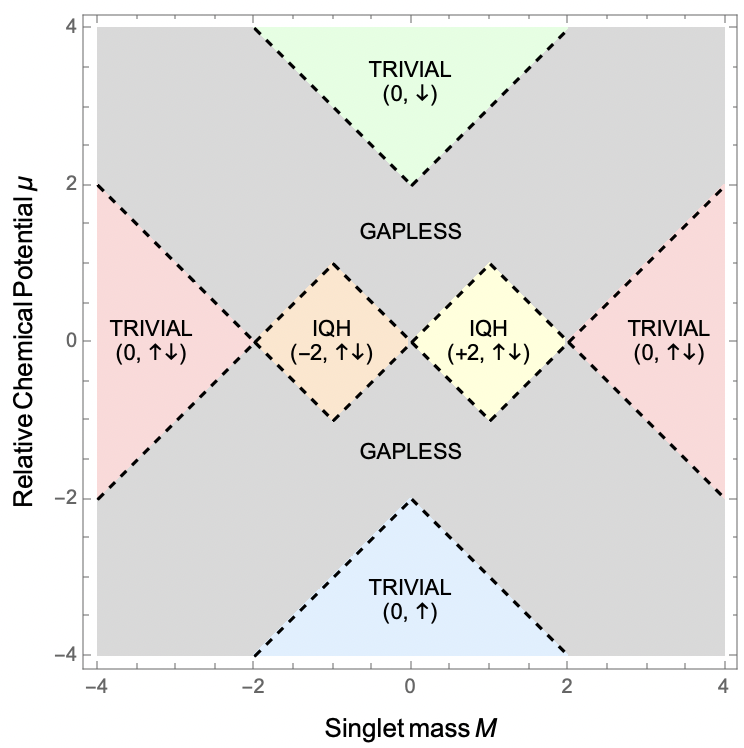}\hspace{5em}
\includegraphics[width = 0.35\linewidth]{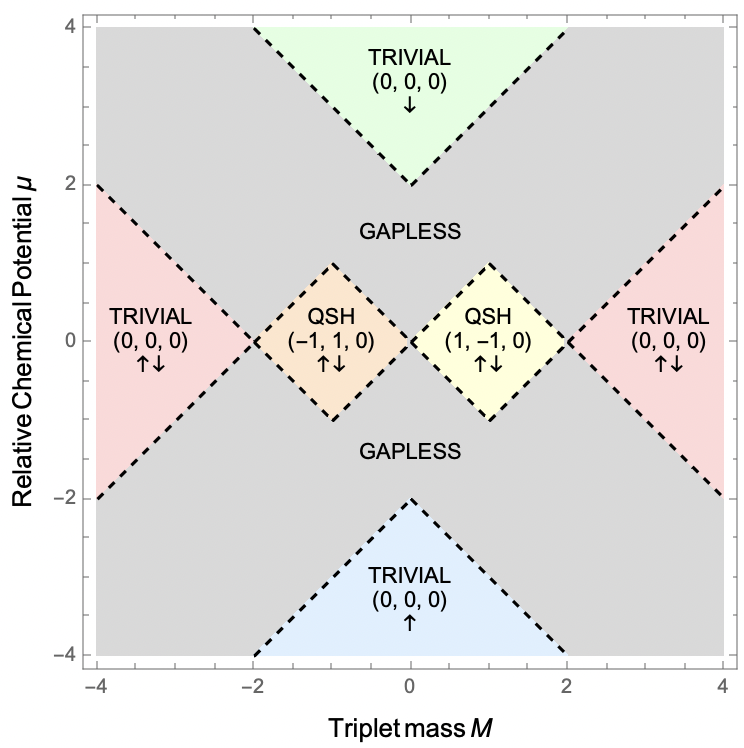}

\caption{The phase diagram of  $N_f=2$ Wilson fermions coupled to a $\U(1)$ gauge field for singlet (left) and triplet (right) mass configurations at weak coupling. In the left figure, the pairs are labeled as $(c_1[b], \expval{s})$. In the right figure, the triplets of Chern numbers are labeled as $(c_\uparrow, c_\downarrow, c_\text{tot})$. Here, ``$\uparrow\downarrow$" denotes zero average spin. The diagram highlights the existence of IQH phases for singlet masses and QSH phases for triplet masses. 
This plot was produced on a $16\times 16$ lattice using the analytical solution~\cite{PRD}, which allows access to large system sizes.\label{fig:PhaseDiag} Details on deriving the phase diagram may be found in Ref.~\cite{PRD}.}
\end{figure*}
\emph{Topological phases and time-reversal symmetry of lattice fermions}--. In this section, we explain the relation between time-reversal symmetry and topological phases characterized by a Chern number and highlight an apparent conflict between the continuum and the lattice. Since the gauge-kinetic terms are time-reversal invariant, it suffices to focus on the fermionic terms coupled to a background $\U(1)$ gauge field.

In the continuum, the theory of a single Dirac fermion coupled to a $\U(1)$ gauge field exhibits the so-called parity ``anomaly",\footnote{This is \textit{not} an anomaly, but rather an explicit breaking. This terminology is simply a convention (see Chap. 8 of~\cite{TongGaugeTheory2018}).} which is manifest in its effective low-energy Chern-Simons description. In other words, since charge conjugation is a UV symmetry and CPT is always a symmetry, both time-reversal and parity are broken explicitly in the continuum. In contrast, the lattice Hamiltonian formulation of a single staggered fermion coupled to a $\U(1)$ gauge field is shown to be time-reversal invariant~\cite{PRD}, which implies that the Chern number of the Berry curvature is vanishing. Therefore, the low-energy limit of staggered fermions is always in a topologically trivial phase with a vanishing Chern-Simons level.

These observations can be rephrased in terms of fermionic doublers. The continuum theory of two Dirac fermions coupled to a $\U(1)$ gauge field is time-reversal invariant provided that the two masses are equal and opposite. On the lattice Hamiltonian side, the staggered fermion formulation reduces the doublers from four to two but does not eliminate them entirely. Due to the staggered mass, these two modes have equal and opposite mass, making the theory time-reversal invariant. 

\begin{table}[H]
\centering
\begin{tabular}{l|c}
\hline
\textbf{Fermion Discretization} & \textbf{Time-reversal broken?} \\
\hline
Continuum $N_f=1$ Dirac     & \cmark\; for any $m$ \\
Continuum $N_f=2$ Dirac     & \xmark\; iff $m_1=-m_2$ \\
Lattice $N_f=1$ Staggered   & \xmark\;  for any $m$\\
Lattice $N_f=1$ Wilson      & \cmark\; for any $(m, R)$\\
Lattice $N_f=2$ Wilson      & \cmark\; iff $(m_1, R_1)=-(m_2, R_2)$ \\
\hline
\end{tabular}
\caption{Summary of time-reversal symmetry breaking.}
\label{tab:TimeReversal}
\end{table}
To construct a time-reversal breaking Hamiltonian of a single doubler-free fermion coupled to a $\U(1)$ gauge field, we use Wilson fermions. The Wilson term removes all doublers in the Hamiltonian formulation and breaks time-reversal symmetry explicitly. This is consistent with well-known results from the Lagrangian formulation on the topological phases of Wilson fermions. Our findings from \cite{PRD} are summarized in Table~\ref{tab:TimeReversal}.

\emph{Topological phases of $N_f=1$ QED$_3$}---. We describe the temporal gauge Hamiltonian and topological phases of a single $(2+1)$D complex two-component Wilson fermion of mass $m$ and Wilson coupling $R$ coupled to a $\U(1)$ gauge field.\footnote{Hamiltonian and Lorentzian $\g$-matrix conventions are in \cite{PRD}.} The Hamiltonian $H = H_f+H_B+H_E$ is ($k=x, y$):
\begin{align}\label{H1}
    H_f &= \frac{1}{2} \sum_{r} \left[\psi^\dagger_{r}\gamma^0\left(i\gamma^k + R \right) \psi_{r+\hat{k}}\sfU_{k}(r) + \textit{h.c.}\right] \no\\
    &\hspace{10em}+ (m+2R)\sum_{r} \psi^\dagger_r \gamma^0 \psi_r,\no\\
    H_E &= \frac{e^2}{2}\sum_{k,r} \sfE_k(r)^2,\hspace{3em}H_B= - \frac{1}{e^2}\sum_r\cos\sfB(r). 
\end{align}
Physical states satisfy Gauss' law at each vertex: $\sfG(r)\ket{\psi} = 0\text{ for all }r$, where $\sfG(r) = \Delta^{-}_{k,r}\sfE_k(r) - \cQ(r)$, $\Delta^{-}_{k,r}$ is the negative-difference operator, $\cQ(r) = \psi(r)^\dagger\psi(r)$ is the charge-density operator, and $\sfE_k(r)$ is the electric field operator that satisfies $[\sfE_k(r), \sfU_\ell(r')] = e\delta_{k,\ell}\delta_{r,r'}\sfU_k(r)$ ($e$ is the gauge coupling). Averaging over the group of all Gauss' law operators, we define a projector $\amsmathbb{P}$ onto the physical Hilbert space $\cH_\text{phys}$, which commutes with the Hamiltonian $H$. A state $\ket{\psi}\in\cH_\text{phys}$ iff $\amsmathbb{P}\ket{\psi} = \ket{\psi}$. 

Observe that $H_f+H_B$ commutes with the non-contractible Wilson lines, $\cW_x = \prod_{x} \sfU_x(x, 0)$ and $\cW_y = \prod_y \sfU_y(0, y)$, which generate an approximate $\U(1)\times \U(1)$ flux-symmetry at weak coupling when $H_E$ is small relative to the mass-gap $\Delta\ne 0$.\footnote{We assume that the mass $M=m+2R$ is chosen so that we are \textit{not} at a gapless point.} In the thermodynamic limit at weak coupling, the ground state lies in the trivial-flux sector, i.e. $(\cW_x, \cW_y) = (1,1)$ \cite{PRD}. In this sector, we find that $\sfU_k = 1$ for $k=x, y$ up to gauge-equivalent configurations. Subject to this gauge-background, we solve the (Wilson) fermionic theory, which exhibits non-trivial topological phases. Following this, we uplift this solution to be gauge-invariant using the Gauss-projector $\amsmathbb{P}$, which does not change the spectrum or location of gapless transition points. In terms of the many-body fermionic solution $\ket{f_g}$ subject to the gauge background $\ket{g}=\bigotimes_{k, r}\ket{\sfU_k(r)=1}$, the gauge-invariant solution is $\amsmathbb{P}\ket{f_g, g}$, which has the same energy as $\ket{f_g,g}$ since $[H,\amsmathbb{P}] = 0$ (see the penultimate section and \cite{PRD}). In this state, the first Chern number of the Berry curvature $b$ on the Brillouin zone $\cB$ is \cite{PRD}
\begin{align}
    C & = \int_\cB \frac{b}{2\pi}=\begin{cases}
        -1,&\text{ if }M\in(-2,0)\\
        +1,&\text{ if }M\in(0,+2)\\
        0,&\text{ otherwise }
    \end{cases}
\end{align}

Fig.~\ref{fig:ChNo} is consistent with the proposal of \cite{Sen:2020srn, Mazza:2011kf} in the absence of gauge fields. In \cite{PRD}, we study the Chern-Simons level (derived from the lattice Lagrangian \cite{Sen:2020srn}) in the limit where the temporal lattice-spacing $a_0\rightarrow 0$. We find full agreement with Fig.~\ref{fig:ChNo}, including the locations of the transitions at $M=m+2R=0, \pm 2$ (we set $R=1$). This derivation fills the missing link between the Hamiltonian in Eq.~\eqref{H1} and the Chern insulator Hamiltonian alluded to in \cite{Sen:2020srn} in the absence of gauge fields. We show that it is consistent to turn off gauge fields in the $\cW_k$-invariant vacuum \cite{PRD}.

\emph{Topological phases of $N_f=2$ QED$_3$}---. We consider the topological phase diagram of the two-flavor theory as a function of the masses, both at zero and at finite density. At zero chemical potential, the Hamiltonian with $N_f = 2$ flavors coupled to a $\U(1)$ gauge-field and relative chemical potential $\mu$ is
\begin{align}
    H_f = &\frac{1}{2} \sum_{r} \left[\psi_a(r)^\dagger\gamma^0\left(i\gamma^k + R \right) \psi_a(r+\hat{k})\sfU_{k}(r) + h.c. \right]\no\\
    & +\sum_{r} M_{a} \psi_a(r)^\dagger \gamma^0 \psi_a(r)+\mu(\psi_1^\dagger\psi_1 - \psi_2^\dagger\psi_2),
\end{align}
where we sum over $a =\; \uparrow,\; \downarrow$. In the second line above, we have implicitly included the dependence on the Wilson couplings inside the shifted mass-matrix $M_{a} = m_{a} + 2 R$. We will focus on the following special cases: $M_1=M_2=M$ (only a singlet mass) and $M_1=-M_2=M$ (only a triplet mass). We focus on the theory at half-filling, defined by $\frac{1}{4L^2} \sum_r \psi_a(r)^\dagger\psi_a(r) = \half$. Since the arguments made in favor of the gauge choice $\sfU_k=1$ do not depend on the number of flavors, we may analytically compute the Chern number and observables like the average occupation fractions \cite{PRD}, which serve as a diagnostic for metal-insulator transitions
\begin{align}
    f_a(\mu, M) & = \frac{1}{2L^2}\sum_{k\in\cB}\expval{\psi_a(k)^\dagger\psi_a(k)}_{\mu, M} 
\end{align}
where $\sum_a f_a(\mu, M) = 1$. 

Our investigation reveals a rich finite-density phase diagram, hosting a variety of novel topological phases. The full weak-coupling infrared phase diagram in Fig.~\ref{fig:PhaseDiag} was obtained by evaluating the occupation fractions and Chern numbers over the entire $(M, \mu)$-plane for both the singlet and triplet cases. In the insulating (i.e. gapped) phases, we find that $(f_\uparrow, f_\downarrow)\in \{(1,0), (\half, \half),(0, 1)\}$, which we label as $\braket{s}\in\{\uparrow, \uparrow\downarrow, \downarrow\}$ in Fig.~\ref{fig:PhaseDiag} respectively. The gapped phases may be topological, i.e. integer quantum Hall (IQH) phases for the singlet case and quantum spin Hall (QSH) phases,  or trivially gapped but characterized by $\braket{s}$. In addition, the phase diagram also hosts gapless metallic phases. Each metal-insulator phase transition is found to be second-order \cite{PRD}. 

We emphasize that simulating such novel topological phases is firmly beyond the reach of classical methods due to the fermion sign problem that plagues QED$_3$ \textit{and} the exponential computational complexity of simulating fermions. Our results from exact diagonalization pave the way for near-term quantum simulations of such topological phases, which serve as an experimental verification of our predictions.

\emph{A path to Wilson fermion quantum simulations with gauge fields}---. So far, we have shown that lattice Hamiltonian theories with Wilson fermions capture a variety of topological phases while staggered fermions fail to do so. However, staggered fermions have received a lot of attention as far as implementation on quantum computing platforms is concerned, while similar experiments and quantum simulations with Wilson fermions are not as widespread. 

In this section, we provide numerical results obtained via exact diagonalization (ED) on $2\times2$ systems with $\integers_N$ gauge fields, which serve as a starting point for a deeper study of quantum simulations with Wilson fermions. This also provides an independent check on the results developed so far. 

When we include gauge fields, there are a number of considerations. First, the gauge fields must be truncated from $\U(1)$ to $\integers_N$. Second, the gauge-invariant physical states that carry both fermionic and gauge labels must satisfy Gauss' law. Third, the symmetries imposed prior to introducing the gauge fields must be compatible with the gauge symmetry, i.e. the Gauss-law projector and the symmetry-generators must share simultaneous eigenstates, as shown explicitly for the weakly-coupling flux-symmetry generated by $\cW_k$ for both the $\integers_N$ and $\U(1)$ cases \cite{PRD}. Since our theoretical arguments indicate that the lowest $\cW_k$-invariant state is independent of the gauge field values, we must observe the topological transitions from ED even for $N=2$ as proof of concept. This offers the possibility to realize such topological phases on near-term quantum computing platforms (such as superconducting devices~\cite{Cochran:2024rwe}, Rydberg atoms~\cite{gonzalez2024observation}, {trapped} ion~\cite{Meth:2023wzd}, dipolar molecules~\cite{luo2020framework} and fermion-pair registers~\cite{sun2023quantum}) via a spin representation for the truncated $\U(1)$ gauge group.

\begin{figure}[t]
    \centering
    \includegraphics[width=.95\linewidth]{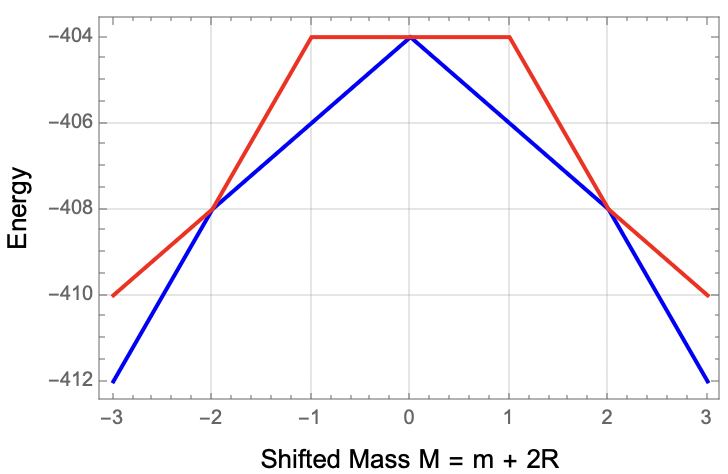}
    \caption{ The 
    level crossings obtained by exact diagonalization on a $2\times2$ lattice for a single Wilson fermion with shifted mass $M=m+2R$ coupled to $\integers_2$ gauge fields at $e^2 = 0.01$. The blue line denotes the ground state and the red line denotes the first excited state, both in the trivial-flux sector with $(\cW_x,\cW_y) = (1, 1)$. Details on the projection to the trivial-flux sector and the unprojected plot may be found in Ref.~\cite{PRD}.}
    \label{fig:ED2x2}
\end{figure}

In Fig.~\ref{fig:ED2x2}, we have displayed a representative sample of the results obtained via ED for the case of $N_f=1$ Wilson fermion with $\integers_2$ gauge fields. Note that the blue and red curves in Fig.~\ref{fig:ED2x2} agree precisely with the analytically predicted gaplessness at $M=0, \pm 2$. We emphasize that the energy levels displayed in Fig.~\ref{fig:ED2x2} are explicitly ensured to be $\cW_k$-invariant by the projectors $\half(\mathbb{1} + \cW_k)$, which commute both with $\amsmathbb{P}$ and $H$. 

As shown in the companion paper \cite{PRD}, the level crossings originate from the Wilson fermions and are robust under turning on $\U(1)$ gauge fields. This robustness of the topological phase transitions persists despite severe gauge group truncation and finite-size effects, which is further confirmed in Fig.~\ref{fig:ChNo} by comparing the Chern number obtained from ED with dynamical $\integers_2$ gauge fields at $e^2=0.01$ with the analytical result obtained for arbitrary lattice size $L$ and $N$ values. The ED data was obtained by computing the many-body Chern number using the Fukui-Hatsugai-Suzuki algorithm by imposing twisted boundary conditions \cite{Fukui:2005wr}. Given this excellent agreement, we conclude that including fully dynamical gauge fields is consistent with the physical picture that has been developed so far. A thorough numerical investigation as a function of $e^2$ and $m$ of the spectrum and a variety of topological observables (such as current correlators) for both $N_f=1\text{ and } 2$ may be found in \cite{PRD}. 

\textit{Future directions---.} In future works, we will examine in further depth the fate of topological phases at strong coupling, and the crossover into the confining regime. In this work, we have focused our attention to the already-rich weakly coupled phase diagram. It is worth emphasizing that the $\cW_k$-symmetry exists only at weak-coupling, i.e. it commutes with $H_f$ and $H_B$, but not the electric term, which is what allows us to extend our analysis to weakly coupled QED$_3$ with $e^2\ll\Delta$, where $\Delta$ is the mass gap. While the topological phases persist when the gauge coupling is sufficiently small compared to the mass gap, there are indeed points in the parameter space where the theory is gapless and $\Delta\rightarrow 0$. In this limit, there is no notion of ``weak coupling" and the theory is necessarily strongly coupled. Therefore, we are required to include the effects of the electric term. Sufficiently far from gapless points where $e^2\ll\Delta\neq 0$, the topological phases indeed persist \cite{PRD}. Given the strongly coupled nature of the gapless points, it is difficult to precisely predict the values of the transition points. We hope that simulations on quantum computers can provide further insights into such phase transitions. It would be most interesting to study the full phase diagram as a function of the gauge coupling, the masses and the chemical potential, especially at the gapless points. Such regimes could be challenging for current classical algorithms due to strong correlations and sign problems~\cite{troyer2005computational}, which motivates the development of quantum simulations. Furthermore, Lagrangian lattice QCD$_4$ with Wilson fermions has been shown to exhibit so-called Aoki phases in the gauge coupling-mass plane \cite{Aoki:1983qi,Aoki:1995ft,Sharpe:1998xm,Magnifico:2018wek,Magnifico:2019ulp}, which maps out the boundary of the topological phases at finite coupling. It would be fruitful to explore such phases in the context of both lattice QED$_3$, and lattice QCD$_4$ at finite density.

Our analysis provides a method for detecting phase transitions with quantum algorithms such as variational quantum eigensolvers~\cite{kandala2017hardware} and quantum phase estimation~\cite{kitaev1995quantum} in the future. While Fig.~\ref{fig:ED2x2} focuses on $N_{f}=1$ Wilson fermion, the truncation scheme also works for $N_{f}=2$ Wilson fermions. An exciting direction is to design efficient quantum algorithms to realize the rich phase diagram in our theoretical analysis and identify the phase transitions in experiments.

\emph{Conclusion}---. In this work, we have analyzed how infrared topological phases can arise in fermionic theories coupled to $\U(1)$ gauge fields. First, we note that theories of staggered fermions are incompatible with the existence of phases characterized by non-trivial Chern numbers. Second, we find that Wilson fermions can support such phases in the Hamiltonian formulation. This is in full agreement with previous literature that has used the Lagrangian formulation to analyze this problem. Third, we mapped out the phase diagram of the QED$_3$ at half-filling with both one and two flavors at zero and finite density, demonstrating metal-insulator transitions and topological (Chern insulator and quantum spin Hall) phases. We have shown that the vacua that we consider are compatible with Gauss' law, as explicitly checked using exact diagonalization. This provides a foundation for realizing novel topological phases on near-term quantum computers originating from Wilson fermions, which are shown to be robust under weakly coupling with $\U(1)$ gauge fields. Our results reveal rich phases of Wilson fermions coupled to gauge fields and pave the way for quantum simulations of topological phases in $(2+1)$D lattice field theory.

\emph{Acknowledgements}---. The authors would like to gratefully acknowledge conversations with Stefan Kühn, Cristina Diamantini, Pranay Naredi, Syed Muhammad Ali Hassan, Srimoyee Sen, Carsten Urbach, Zhuo Chen, Penghao Zhu, Yugo Onishi, Taige Wang, Eric D'Hoker, Theodore Jacobson and Xiaoliang Qi. The authors gratefully acknowledge the granted access to the Marvin cluster hosted by the University of Bonn. SB is supported by the Mani L. Bhaumik Institute for Theoretical Physics. This project was funded by the Deutsche Forschungsgemeinschaft (DFG, German Research Foundation) as part of the CRC 1639 NuMeriQS -- project no.\ 511713970 and under Germany's Excellence Strategy – Cluster of Excellence Matter and Light for Quantum Computing (ML4Q) EXC 2004/1 – 390534769. This work is supported with funds from the Ministry of Science, Research and Culture of the State of Brandenburg within the Centre for Quantum Technologies and Applications (CQTA). 
\begin{center}
    \includegraphics[width = 0.08\textwidth]{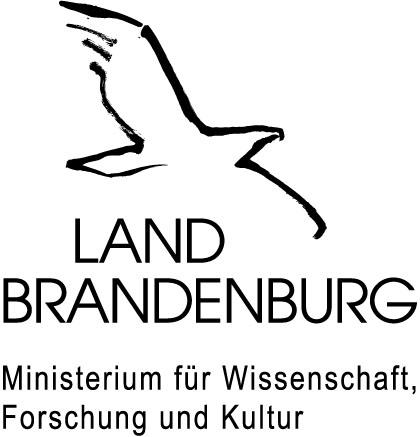}
\end{center}
This work is funded by the European Union’s Horizon Europe Frame-work Programme (HORIZON) under the ERA Chair scheme with grant agreement no. 101087126. This work is part of the Quantum Computing for High-Energy Physics (QC4HEP) working group.

\bibliographystyle{apsrev4-1}
\bibliography{main_PRL}

\end{document}